\documentclass[aps]{revtex4}

\usepackage{amsmath,amssymb,amsfonts}
\usepackage{graphicx}
\usepackage{color}
\allowdisplaybreaks
\newcommand{\be}{\begin{equation}}
\newcommand{\ee}{\end{equation}}
\newcommand{\bea}{\begin{eqnarray}}
\newcommand{\eea}{\end{eqnarray}}
\newcommand{\bm}{\bibitem}

\newcommand{\gm}{\gamma}
\newcommand{\gf}{\gamma_5}
\newcommand{\gz}{\gamma_0}
\newcommand{\Gm}{\Gamma}
\newcommand{\ep}{\epsilon}
\newcommand{\de}{\delta}
\newcommand{\De}{\Delta}
\newcommand{\om}{\omega}

\newcommand{\tht}{\theta}

\newcommand{\lm}{\lambda}
\newcommand{\Lm}{\Lambda}

\newcommand{\Sg}{\Sigma}

\newcommand{\oS}{\overline{S}}
\newcommand{\oSg}{\overline{\Sigma}}

\newcommand{\oom}{\overline{\om}}

\newcommand{\ps}{p \!\!\! /}

\newcommand{\ks}{k \!\!\! /}
\newcommand{\us}{u \!\!\! /}

\newcommand{\vk}{\vec k}
\newcommand{\vp}{\vec p}

\newcommand{\la}{\langle}
\newcommand{\ra}{\rangle}

\newcommand{\bsg}{\boldsymbol{\Sigma}}
\newcommand{\bk}{\boldsymbol{k}}
\newcommand{\bp}{\boldsymbol{p}}
\newcommand{\bS}{\boldsymbol{S}}
\newcommand{\rw}{\rightarrow}

\newcommand{\F}{F_\pi}
\begin{document}

\setcounter{page}{1}

\title{Relativistic spectral function of nucleon in hot nuclear matter}

\author{Sabyasachi \surname{Ghosh}}
\email{sabyasachighosh@veccal.ernet.in}
\affiliation{Variable Energy Cyclotron Centre, 1/AF, Bidhannagar,
Kolkata, 700064, India}
\author{S. \surname{Mallik}}
\email{mallik@theory.saha.ernet.in}
\affiliation{Theory Division, Saha Institute of Nuclear Physics, 1/AF
 Bidhannagar, Kolkata 700064, India}  
\author{Sourav \surname{Sarkar}}
\email{sourav@veccal.ernet.in}  
\affiliation{Variable Energy Cyclotron Centre, 1/AF, Bidhannagar,
Kolkata, 700064, India}

\date{\today}

\begin{abstract}
We present a simple calculation of the nucleon self-energy in nuclear matter 
at finite temperature in a relativistic framework, using the real time thermal 
field theory. The imaginary parts of one-loop graphs are identified with 
discontinuities across the unitary and the Landau cuts. We find that in general 
both the cuts contribute significantly to the spectral function in the region of
(virtual) nucleon mass usually considered, even though the unitary cut is 
ignored in the literature. Also our relativistic spectral function
differs from the one in non-relativistic approximation, used
in some earlier calculations.
\end{abstract}


\maketitle

Heavy ion collisions provide an opportunity to investigate particle
propagation through strongly interacting media. However, only the vector
mesons, particularly the $\rho$, can at present be studied directly
by detecting dileptons, into which they decay in the hot, dense media. The  
media created by these collisions consist, in general, not only of mesons, 
but also of nucleons. Thus the effects of both mesons and nucleons on the 
vector meson spectral functions have been extensively studied in the 
literature \cite{Rapp1}. For a more complete picture, the self-energy of 
nucleon itself need be investigated \cite{Leutwyler,Hees}. The nucleon 
self-energy function also determines the equation of state of nuclear 
matter \cite{Haar}.

In this work, we find the one loop corrections to the nucleon propagator at 
finite temperature and nucleon chemical potential in the real time formulation of 
the thermal field theory \cite{Niemi}. However, to keep an eventual contact with 
the fundamental $QCD$ theory, we do not start directly with the propagator, namely 
the two point function of the nucleon field. Instead, we consider the same of the 
nucleon current $\eta(x)$ \cite{Ioffe,Chung}, which in {\it vacuum} is
\be
i\int d^4xe^{ip\cdot x}\la  0| T \eta (x) \overline\eta(0)|0 \ra\,.
\ee
Here $\eta(x)$ is built out of three quark fields, so as to have the quantum numbers of 
the nucleon. We denote its matrix element between vacuum and nucleon state as
\be
\la  0|  \eta (x) |N (p) \ra = \lm u(p) e^{ip\cdot x}\,,
\ee
where $u(p)$ is the Dirac spinor of the nucleon and the parameter $\lm $ denotes the 
coupling of $\eta(x)$ with the nucleon \cite{footnote1}.

In {\it nuclear matter}, the vacuum expectation value in Eq. (1) must be replaced by 
the ensemble average. Denoting the time-ordered product of the operators by $O(x)$, it means
\be
\la 0| O(x) |0  \ra ~\rw ~Tr\,[e^{-\beta(H-\mu N)}O(x)]/Z,~~~~~~Z=Tr\,[e^{-\beta(H-\mu N)}]\,,
\ee
where $H$ and $N$ are the Hamiltonian and the nucleon number operator of the system , 
$\beta$ is inverse temperature and $\mu$ is the chemical potential corresponding to $N$.
In the real time version, every two point function, including the self-energy we calculate 
below, assumes the form of a $2 \times 2$ matrix. But each of these matrices may be 
diagonalised, when it is given essentially by single analytic function, that determines 
completely the dynamics of the corresponding two-point function \cite{Kobes}. As this 
function is simply related to any one, say the $11$-component of the matrix, we need 
calculate only this component of the self-energy matrix. 

The $11$-component of a free, thermal matrix propagator for a particle is a sum of its 
vacuum propagator and a term depending on the (on-shell) distribution functions of 
like-particles in the medium through which it propagates. The latter term has a 
universal form, depending only on the bosonic or the fermionic character of the 
particle \cite{MS}. Anticipating the propagators for pion, nucleon and
$\De(1237)$ to appear in our calculation, we begin by writing their 11-components.
These particles are represented respectively by scalar $(\phi(x))$, Dirac
$(\psi(x))$ and Rarita-Schwinger $(\De_\mu(x))$ fields \cite{Rarita}. For the bosonic
propagator, it is
\be
D^{11}(k_0,\vk, m_\pi)=\De (k,m_\pi) +2\pi i n(\om)\de(k^2-m_\pi^2)\,,
\ee
where $\De$ and $n$ are the vacuum propagator of the scalar field and its
equilibrium particle distribution function,
\be
\De (k,m_\pi)=\frac{-1}{k^2-m_\pi^2+i\eta}\,,~~~~n(\om)=\frac{1}{e^{\beta \om} -1}\,,~~~~
\om=\sqrt{\vk^2+m_\pi^2}\,.
\ee
For the fermionic propagators, we introduce
\be
E^{11}(p_0,\vp,m)=\De (p,m) -2\pi i N(p_0)\de (p^2-m^2)\,,
\ee
where $N(p_0)$ consists of distribution functions for the particle and anti-particle,
\be
N(p_0)=n_+ (\om')\tht (p_0) +n_- (\om') \tht (-p_0)\,,~~~~
n_{\pm}(\om')=\frac{1}{e^{\beta(\om'\mp\mu)}+1}\,,~~~~\om'=\sqrt{\bp^2+m^2}
\ee
in terms of which the $11$-components of spin $\frac{1}{2}$ and
$\frac{3}{2}$ propagators may be written respectively as 
\bea
&& S^{11}(p_0,\vp)=(\ps+m_N)E^{11}(p_0,\vp,m_N)\,,\\
&& S^{11}_{\mu\nu} (p_0,\vp)=(\ps+m_\De)\left\{-g_{\mu\nu}+\frac{2}{3m_\De^2}
p_\mu p_\nu +\frac{1}{3}\gm_\mu \gm_\nu +\frac{1}{3m_\De}(\gm_\mu p_\nu-\gm_\nu
p_\mu)\right\} E^{11} (p_0,\vp,m_\De)
\eea

\begin{figure}
\includegraphics[width=8cm]{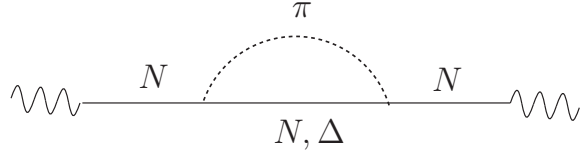}
\caption{One-loop graphs for the two-point function contributing to
self-energy of the nucleon}
\end{figure}   

The spectral function of nucleon is obtained from the Dyson equation, giving the 
complete propagator $\bS'$ in terms of the free propagator $\bS$  and self-energy $\bsg$ ,
\be
\bS' =\bS - \bS \bsg \bS'
\ee
where each element is a $2 \times 2$  matrix in the thermal indices (besides being 
$4 \times 4$ matrices in Dirac space). As already stated, they can all be diagonalised to 
get the respective analytic functions, denoted by a bar, satisfying
\be
\overline{S'} = \overline{S} - \overline{S}~ \overline{\Sg}~ \overline{S'}
\ee
which may be readily solved as usual. The self-energy function is related to
the $11$-component of the corresponding matrix by \cite{Kobes},
\bea
{\rm Re}\,\oSg(p) &=&{\rm Re}\,\Sg^{11}(p)\nonumber\\
{\rm Im}\,\oSg(p) &=&\coth[\beta (p_0-\mu)/2]{\rm Im}\,\Sg^{11}(p)
\eea

The free propagator $ \overline{S}$ turns out to be the same as in vacuum,
\be
\oS (p)=\frac{- (\ps+ m_N) }{p^{2}-m_N^2+i\eta}
\ee
The calculation simplifies if we take $ \bp =0$ . Also restricting to the anti-nucleon 
pole in Eq.(13), it becomes
\be
\oS(p_0) =\frac{(1+\gz)}{2} \frac{-1}{p_0-m_N+i\eta}\,,~~~~
\ee
Decomposing $\oSg$ and $\oS'$ in Dirac space,
\be
\oSg~=~\Sg_s+\gz\Sg_v\,,~~~~\oS'~=~S'_s+\gz S'_v \,,
\ee
it  follows from Dyson equation that $S'_s  = S'_v$. Then
letting  $\Sg = \Sg_s+\Sg_v$, we get the complete propagator as
\be
\oS' (p_0)=\frac{(1+\gz)}{2}\frac{-1}{p_0-m_N-\Sg}
\ee
giving the spectral function
\bea
A_N(p_0)=\frac{-{\rm Im}\Sg}{(p_0-m_N-{\rm Re}\Sg)^2+({\rm Im}\Sg)^2}
\eea

The graphs which we wish to evaluate are shown in Fig. 1. Because we are interested in
finding the nucleon self-energy, we retain only the graphs which couple the nucleon 
current to the nucleon. Also we include $\De(1237)$ resonance besides the
nucleon in the intermediate state. We shall comment later on the contribution of higher 
resonances.

\begin{figure}
\includegraphics[scale=0.35]{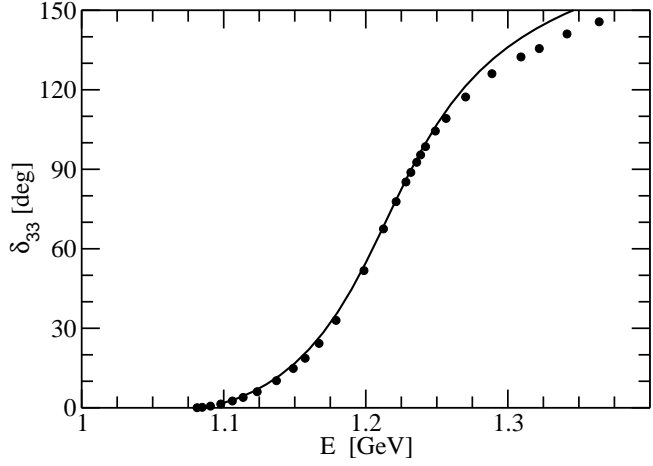}
\caption{Graph of phase shift for $P_{33}$ partial wave in $\pi N$ scattering from our 
model compared with experiment~\cite{Koch} (solid circles)}
\end{figure}   

The vertices appearing in Fig.1 may be obtained from chiral perturbation theory 
\cite{Gasser,Ecker,Becher}. The appropriate field variable for pion in the effective 
theory is not $\phi (x)$ introduced earlier, but the $SU(2)$ valued matrix field 
$u(x)$ related to $\phi (x)$ by $ u(x)=\exp(i\vec{\tau}\cdot\vec{\phi}/2\F)$,
where $\F =93 $ MeV, the so-called pion decay constant. The effective Lagrangians 
are \cite{footnote2},
\bea
{\cal L}_{\pi NN}&=&\frac{1}{2}g_{A}\overline\psi \us \gf\psi =
-\frac{g_{A}}{\F}\overline\psi \gm_{\mu}
\gf \tau^{a}\psi\, \partial^{\mu}\phi_{a} + \cdots\\
{\cal L}_{\pi N\De}&=&\frac{g_\De}{\sqrt{2}}\overline\psi_a (u^\mu)^c_b
\De_\mu^{abd}\ep_{cd} + h.c. =
\frac{g_\De}{\F}\overline{p}\,\partial^\mu\pi^- \De_{\mu}^{++} +\cdots
\eea
where $u_\mu=i(u^{\dagger}\partial_\mu u-u\partial_\mu u^{\dagger}),\,
a,b\cdots =1,2 $ and $\ep_{12}=-\ep_{21}=1$, etc. The coupling constants
$g_A$ and $g_\De$ are to be determined phenomenologically. As is well-known
\cite{Korpa,Weinhold}, such a model requires form factors at the vertices, which 
we take in the Lorentz invariant form as
\be 
F(p,k)=\frac{\Lm^2}{\Lm^2+ (p\cdot k/m_N)^2 -k^2}
\ee
where $p$ and $k$ are the four-momenta of nucleon and pion at the vertices
and $\Lm$ is essentially a cut-off on these momenta.

We first check this model with experimental data on $\pi N$ scattering. The
interaction (19) allows us to calculate the decay width of $\De \rightarrow N+\pi$ as 
a function of its energy as 
\be
\Gm (E) = \frac{1}{24\pi} \left(\frac{g_\De}{\F}\right)^2 F^2(E)
|\bp|^3\frac{(E+m_N)^2-m_\pi^2}{E^2}
\ee
Here $\bp$ is the three-momentum in the centre-of-mass of $\pi N$ system,
\be
\bp^2=\frac{\{ E^2-(m_N+m_\pi)^2\} \{E^2-(m_N-m_\pi)^2\}}{4E^2}. 
\ee
In this kinematic configuration, the form factor becomes
\be
F(|\bp|)=\frac{\Lm^2}{\Lm^2 +(|\bp|E/m_N)^2}
\ee
The pion-nucleon partial wave $f$ in the $P_{33}$ channel may now be written in 
the form
\be
f(E)\sim \frac{1}{E^2-m^2_\De +im_\De\Gm (E)}
\ee
We take the resonance parameters at the pole position, $m_\De =1210$ MeV and
$\Gm (m_\De)=100$ MeV \cite{Data}. Taking $g_\De =2.2$ and $\Lm=400$ MeV
\cite{Weinhold}, we can satisfy Eq.(21) and also achieve reasonable agreement of 
the phase shift $\de_{33}$ computed from Eq.(24) with experiment~\cite{Koch} (Fig. 2). Also we 
take $g_A=1.26$ \cite{Weinberg} and the same form factor at the $\pi N N$ vertex.    

\begin{figure}
\includegraphics[width=18cm]{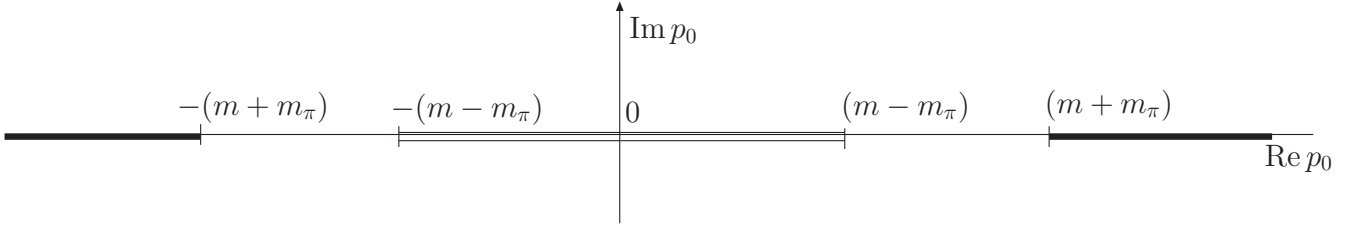}
\caption{Branch cuts of self-energy function in $p_0$ plane for $\bp =0$.}
\end{figure}

We now evaluate the self-energies from graphs of Fig. 1. The $11$-component
of each of the loops has the general form
\be
\Sg^{11}(p_0,\bp)=i\int\frac{d^4k}{(2\pi)^4}R(p,k)D^{11}(k)E^{11}(p-k)
\ee
where $R(p,k)$ includes the form factors and the factor from interaction Lagrangian at 
the vertices, as well as the spin-dependent factor in the propagator, all of which we 
shall write below explicitly. Inserting propagators from Eqs.(4) and (6) in Eq.(25), 
we get
\bea
\Sg^{11} (p_0,\bp)=&& i\int\frac{d^4k}{(2\pi)^4}\frac{R(k_0)}{(k^2-m_\pi^2+i\ep)
\{(p-k)^2-m^2 +i\ep\}}\nonumber\\
&&-\int \frac{d^4k}{(2\pi)^3}R(k_0)
\left\{\frac{N(p_0-k_0)\de((p-k)^2-m^2)}{k^2-m_\pi^2 +i\ep}
-\frac{n(k_0)\de((k^2-m_\pi^2)}{(p-k)^2-m^2 +i\ep}\right\}\nonumber\\
&& +i\int \frac{d^4k}{(2\pi)^2}R(k_0) n(k_0)N(p_0-k_0)\de (k^2-m_\pi^2)\de
((p-k)^2-m^2)
\eea
where $m$ denotes mass of baryon ($m_N$ or $m_\De$) in the loop and 
we show the dependence of $R$ on $k_0$ only, suppressing other variables for
brevity. Here the first term refers to vacuum. The second and the third terms are
medium dependent, with distribution functions appearing respectively linearly and 
quadratically. Observe that the third term is purely imaginary.

\begin{figure}
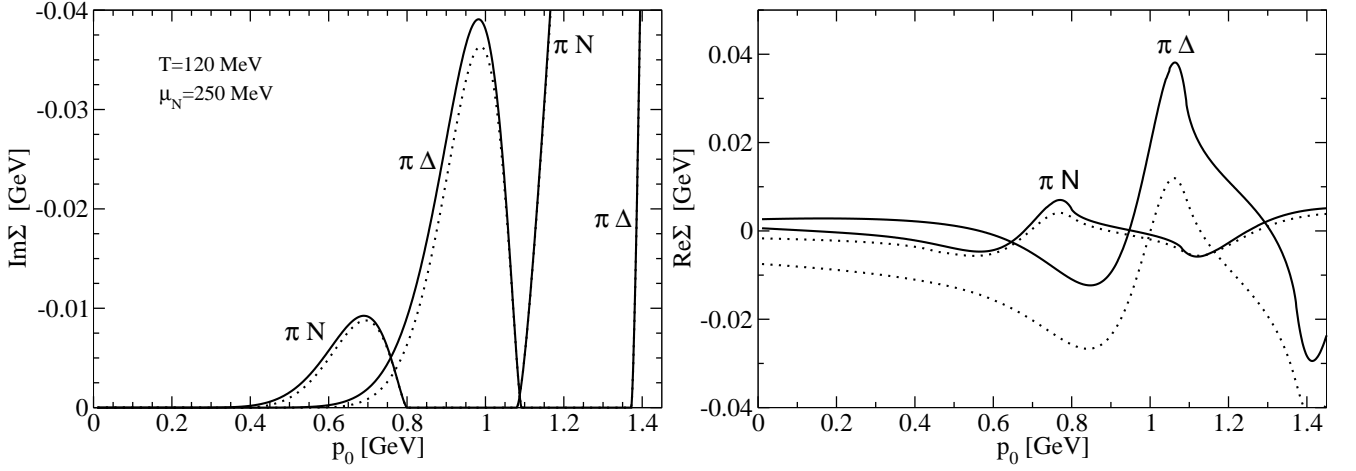

\includegraphics[scale=0.35]{n_im_sig.eps}
\includegraphics[scale=0.35]{n_re_sig.eps}
\caption{Imaginary (left panel) and real (right panel) parts of self-energy from 
$\pi N$ and $\pi \De$ loops. Solid curves represent the results of our calculation
(relativistic, including unitary cuts). Dotted curves result from non-relativistic 
approximation.}  
\end{figure}

\begin{figure}
\includegraphics[scale=0.35]{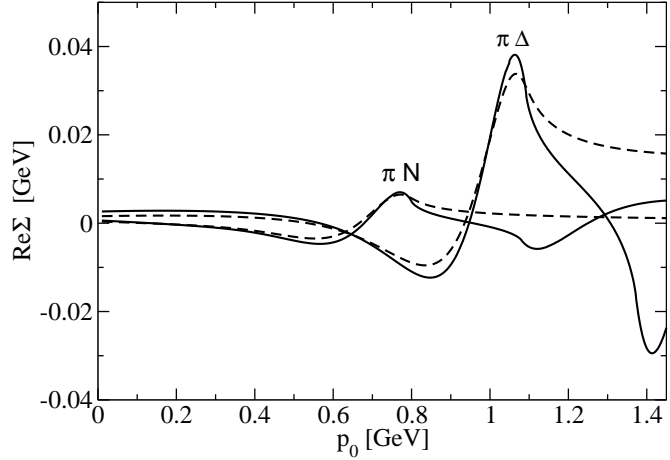}
\caption{Real part of nucleon self-energy with (solid) and without (dashed)
 contribution from the unitary cut.} 
\end{figure}

Carrying out the $k_0$ integration in all the three terms of Eq. (26), we can easily 
find the real and the imaginary parts of $\Sg^{11}$. Though ${\rm Im}\Sg^{11}$ 
contains quadratic terms in distribution functions, ${\rm Im}\oSg$, as defined by 
Eq. (12), turns out to be linear in them \cite{Weldon,Kobes}, 
\bea
&&{\rm Im}\oSg(p_0,\bp)=-\pi\int\frac{d^3\bk}{(2\pi)^3 4\om\om'}\times\nonumber\\
&& [R(k_0=\om)\{(1+n(\om)- n_+(\om'))\de(p_0-\om-\om')
-(n(\om)+n_-(\om'))\de(p_0-\om+\om')\}\nonumber\\
&&  + R(k_0=-\om)\{(n(\om)+ n_+(\om'))\de(p_0+\om-\om')
-(1+n(\om)-n_-(\om')\de(p_0+\om+\om')\} ]
\eea
where $\om$ and $\om'$ are pion and baryon energies,
\be
\om= \sqrt{m_\pi^2+\bk^2} ,~~~~ \om'=\sqrt{m^2+(\bp-\bk)^2} 
\ee
For the real part of $\oSg$, we include only the medium
dependent pieces, which are given by the second term in Eq. (26) alone,
\bea
{\rm Re}\oSg (p_0,\bp)=\int\frac{d^3\bk}{(2\pi)^3 4\om\om'}\left\{
\frac{R_2n+R_3n_+}{p_0+\om-\om'}+\frac{R_1n-R_3n_+}{p_0-\om-\om'}\right.\nonumber\\
\left.-\frac{R_1n+R_4n_-}{p_0-\om+\om'}-\frac{R_2n-R_4n_-}{p_0+\om+\om'}\right\}
\eea
where $R_i,\, (i=1,\cdots ,4)$ denote the function $R(k_0)$ evaluated at
$k_0=\om,\,  -\om,\,  p_0-\om' \, $ and $p_0+\om'$ respectively.

In the following we restrict our evaluation to $\bp=0$, when there will be no angular 
dependence. First consider the imaginary part, giving the cut structure 
\cite{Das,GSM}. The delta-functions in the different terms of Eq. (27) control the 
regions of non-vanishing imaginary parts of $\oSg$, which define the position of the 
branch cuts. As shown in Fig. 3, the first and the fourth terms give rise to the regions, 
$p_0 \geq (m +m_\pi)$ and $p_0 \leq -(m +m_\pi)$ respectively, giving the unitary cuts. 
Similarly the second and the third terms lead to the regions, $0\leq  p_0 \leq (m -m_\pi)$ 
and $-(m -m_\pi)\leq  p_0 \leq 0 $, giving the Landau cuts.

\begin{figure}
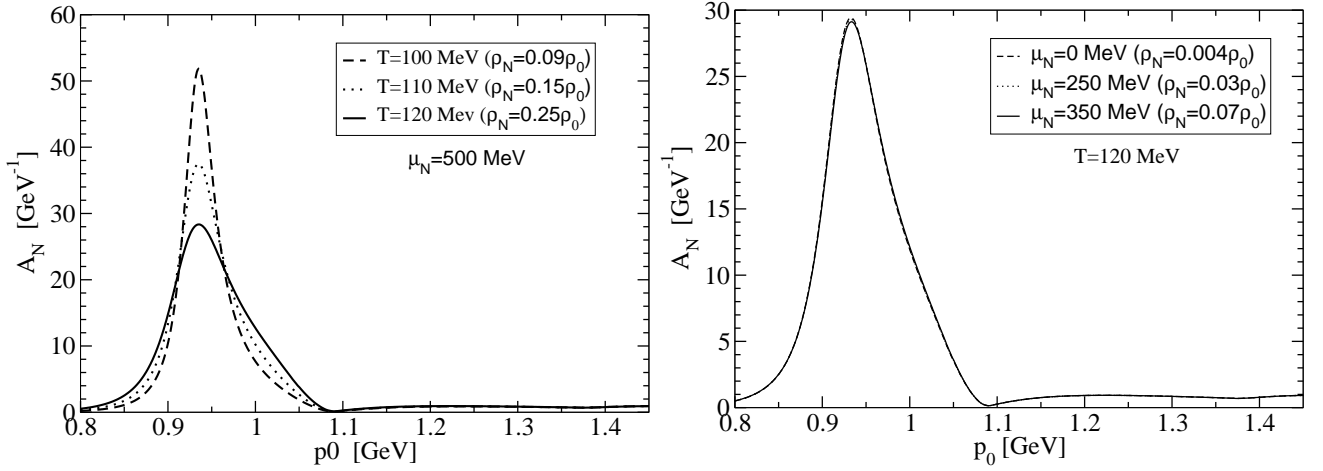

\includegraphics[scale=0.35]{n_spec_mu.eps}
\includegraphics[scale=0.35]{n_spec_T.eps}
\caption{Nucleon spectral function at different temperatures for a fixed
chemical potential (left panel) and at different chemical potentials for a
fixed temperature (right panel)}
\end{figure}  

With $\bp=0$, the value of $|\bk|$, fixed by the $\de$-functions in Eq. (27), is the 
magnitude of three-momentum in the center-of-mass of the pion-baryon system,
\be
|\bk|^2=\frac{\{p_0^2-(m+m_\pi)^2\}\{p_0^2-(m-m_\pi)^2\}}{4p_0^2}
\ee
In this frame, the form factor (23) simplifies to
\be
F(|\bk|)=\frac{\Lm^2}{\Lm^2+\bk^2}
\ee
Let us define the variables,
\be
\oom=\frac{p_0^2+m_\pi^2-m^2}{2p_0},~~~~\oom'=\frac{p_0^2-m_\pi^2+m^2}{2p_0}
\ee
which actually coincide respectively with $\om$ and $\om'$ defined by Eq.(28) on all 
the cuts, except for $\om$ on the Landau cut, where $\om = |\oom|$. Then the imaginary 
parts of $\oSg$ for $p_0\geq 0$ are given by the first and the third terms of
Eq.(27) as
\be
{\rm Im}\oSg (p_0)=-\frac{|\bk|}{8\pi p_0}\left\{ \begin{array}{ll}
\displaystyle R(k_0=\oom) \{1+n(\oom)-n_+(\oom')\}\,, & \mathrm{on\; unitary\; cut}\\
\displaystyle R(k_0=\oom) \{n(|\oom|)+n_+(\oom')\}\,,  &  \mathrm{on\; Landau\; cut}
\end{array}
\right.
\ee

The expression (29) for the real part also simplifies for $\bp =0$ to 
\be
{\rm Re} \oSg (p_0)=-\frac{1}{16\pi^2p_0}P\int_{m_\pi^2}^\infty 
\frac{d\om^2 \sqrt{\om^2-m_\pi^2} h(\om)}{\om\om'(\om^2-\oom^2)}
\ee
where
\be
h(\om)=\om'\{(\om+\oom)R_1-(\om -\oom)R_2\}n
-\om\{(\om'+\oom')R_3n_+ -(\om'-\oom')R_4n_-\}
\ee

\begin{figure}
\includegraphics[scale=0.35]{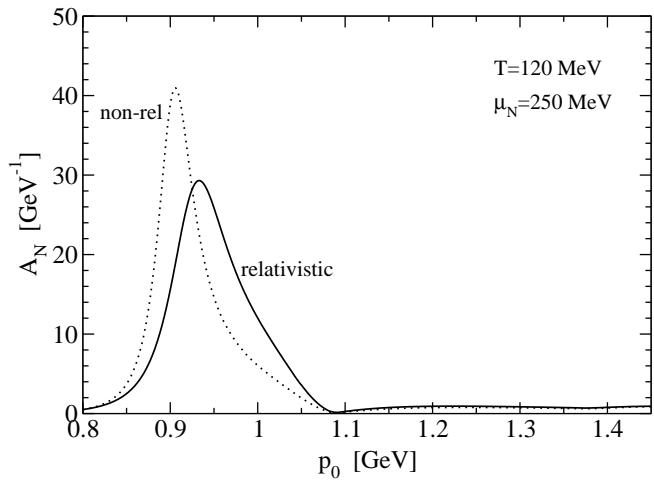}
\caption{Comparison of relativistic spectral function (continuous curve) with 
its non-relativistic limit (dotted curve).}
\end{figure} 

Having carried out the evaluation in terms of the (Dirac) matrix-function 
$R (p,k)$, it remains to write its explicit expressions for the two loops,
\bea
R(p,k)&=&\frac{3}{4}\left(\frac{g_{A}}{F_{\pi}}\right)^2 F^2 (p,k)
\{2k\cdot p \ks -k^2 (\ps + \ks + m_N)\}\,,~~~~~~~~~(\pi N\, \mathrm{loop})\\
R(p,k)&=& \frac{4}{3}\left(\frac{g_\De}{\F}\right)^2 F^2 (p,k)
\left\{-k^2+\frac{(p\cdot k-k^2)^2}{m^2_\De}\right\} (\ps-\ks+m_\De), ~~~~~
(\pi \De\, \mathrm{loop})
\eea

In this calculation, we consider only $N$ and $\De$ in the intermediate
state, as we expect the contributions of higher mass resonances to be small.  
The reason is that on both the cuts all the distribution functions decrease
rapidly with the rise of resonance mass, the only exception being $n(\om)$
for pions on the unitary cut, which does the reverse. However, unlike the 
Landau cut, where all higher mass resonances can contribute, the unitary cut,
for a fixed upper value of the (virtual) nucleon mass, gets contribution from 
only a finite number of resonances. Thus in the present case, where we
restrict $p_0\lesssim 1.5$ GeV, only $N$ and $\De$ can contribute to the unitary
cut.

A question arises in such calculations, whether a non-relativistic approximation 
could reproduce the relativistic results in a quantitative way \cite{Post}. To 
define this approximation, we rewrite $E^{11}(p_0, \bp)$, the spin-independent 
factor in the $11$-component of baryon propagator, given by Eq.(6), as
\be
E^{11}(p_0, \bp) = \frac{-1}{2\om'}\left\{\frac{1-n_+(\om')}{p_0-\om'+i\ep}+
\frac{n_+(\om')}{p_0-\om'-i\ep} -\left(\frac{1-n_-(\om')}{p_0+\om'-i\ep}- 
\frac{n_-(\om')}{p_0+\om'+i\ep}\right)\right\}
\ee
Here the first two terms describe the propagation of baryon and the last two
that of antibaryon. The non-relativistic approximation to this propagator consists 
in retaining only the first two terms above \cite{Fetter,Serot}. This approximation 
in turn gives only the first two terms in Eq.(29) for ${\rm Re} \oSg (p_0,\bp)$. Further
we set $\om'=m$ everywhere for baryon \cite{footnote3}. Note that we approximate 
neither the propagator $D^{11}(k)$ for pion nor its energy-momentum relation.  
 
In presenting the results of numerical evaluation, we consider, along with our 
relativistic framework, also the non-relativistic one as defined above and another where 
the unitary cuts are switched off (by omitting the second and fourth terms in 
Eq. (29)), but relativistic otherwise. We note here 
that the  values of $g_\De$ and $\Lm$ as well as the fit to the experimental data for 
$\de_{33}$ that we found earlier remain unaffected by the non-relativistic approximation. 
Fig. 4 compares the typical behaviour of relativistic results with the non-relativistic 
ones for the imaginary and real parts of self-energy, separately for the two loops
-- we see that only the real part for $\pi\De$ loop differ significantly between the two.
Fig. 5 does this comparison for the real parts between the complete result and the one 
without the unitary cuts, showing significant difference only at higher masses.
(The imaginary parts here are, of course, the same as in Fig. 4 without the
steep lines representing the unitary cuts.) 

Having made these comparisons, we come back to our model in Fig. 6 to draw the nucleon 
spectral function at different values of $T$ and $\mu$, which are realised in heavy-ion 
collisions \cite{Andronic,Cleymans}. As expected, the height of the peak decreases with 
rise of temperature, while it remains about the same within the interval of chemical 
potential considered here \cite{Rapp2}. Finally we again compare in Fig. 7 a typical 
spectral function of our calculation with its non-relativistic limit. 

\begin{figure}
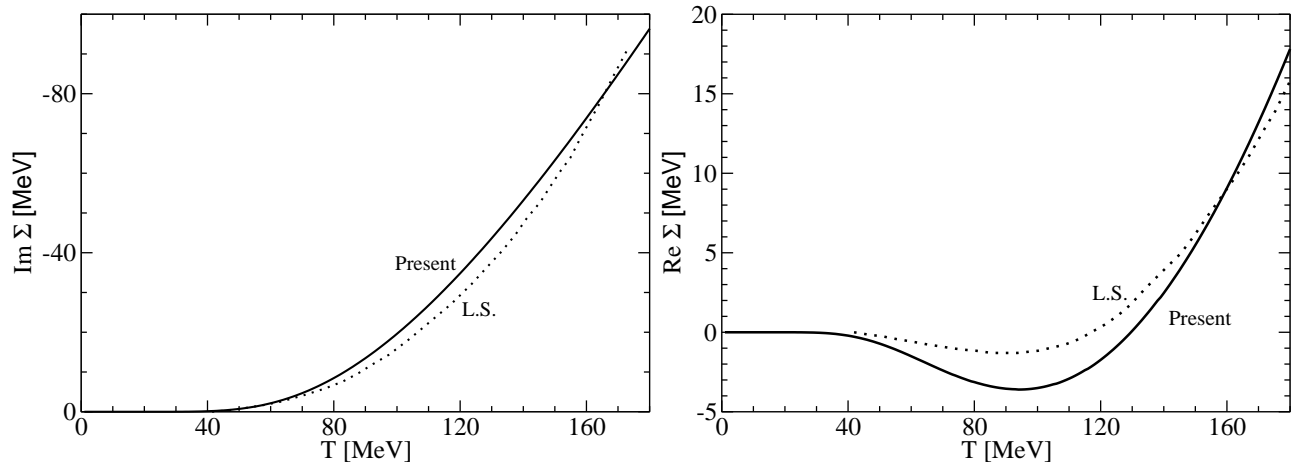

\includegraphics[scale=0.35]{n_im_LS.eps}
\includegraphics[scale=0.35]{n_re_LS.eps}
\caption{ The present evaluation of on-shell imaginary (left panel) and the real 
(right panel) parts of self-energy in pion medium (continuous curve) compared with those 
of Ref. \cite{Leutwyler} (dotted curve.)}
\end{figure} 

We also compare our results with two earlier calculations. Leutwyler and Smilga 
\cite{Leutwyler} use virial expansion to leading order to obtain the self-energy 
for {\it  on-shell} nucleon in pionic medium. In Fig. 8 we compare their results 
for the imaginary and real parts with those from our model, setting nucleon
and $\De$ distribution functions to zero. The good agreement shows that our model is 
realistic, if we recall that they evaluate the virial formula with {\it experimental data}
on $\pi N$ scattering. Hees and Rapp \cite{Hees} calculate the self-energy in the 
imaginary time formulation and takes into account only the Landau cut for the 
imaginary part and correspondingly only the first term in Eq. (29) for the real part. 
However, they take higher mass resonances in the loop, validating their model at higher 
temperatures and chemical potentials. 
 
To conclude, we calculate the self-energy of the nucleon and its spectral function 
in the real time version of the thermal field theory in the relativistic framework. 
The imaginary part of the self-energy is built out of contributions from both Landau 
and unitary cuts from one loop graphs with $\pi N$ and $\pi\De$ intermediate states. 
In contrast to results in the literature, we find the unitary cut from the $\pi\De$ 
loop to contribute significantly in the upper region of (virtual) mass of nucleon 
considered. The nucleon spectral function turns out to be sensitive to non-relativistic 
approximation, establishing the necessity of relativistic treatment for its quantitative 
determination.

\section*{Acknowledgement}

One of us (S.M.) acknowledges support from Department of Science and
Technology, Government of India. We also thank the Referee for his
recommendations leading to improvement of the manuscript.

\end{document}